\documentclass[aps,pra,reprint,groupedaddress,longbibliography]{revtex4-1}
\pdfoutput=1
\usepackage{graphicx,color}
\usepackage{amsmath, amssymb}
\usepackage{longtable}
\usepackage{natbib}

\begin{document}
\title{
New mechanism of kinetic exchange interaction induced by strong magnetic anisotropy
}
\author{Naoya Iwahara}
\author{Liviu F. Chibotaru}
\affiliation{Theory of Nanomaterials Group, 
Katholieke Universiteit Leuven, 
Celestijnenlaan 200F, B-3001 Leuven, Belgium}
\date{\today}

\begin{abstract}
It is well known that the kinetic exchange interaction 
between single-occupied magnetic orbitals (s-s) 
is always antiferromagnetic, 
while between single- and double-occupied orbitals (s-d) is 
always ferromagnetic and much weaker. 
Here we show that the exchange interaction between strongly anisotropic doublets of lanthanides, actinides 
and transition metal ions with unquenched orbital momentum contains a new s-d kinetic contribution
equal in strength with the s-s one. 
In non-collinear magnetic systems, this s-d kinetic mechanism can cause an overall ferromagnetic exchange
interaction which can become very strong for transition metal ions.
These findings are fully confirmed by DFT based analysis of exchange interaction in several Ln$^{3+}$ complexes.
\end{abstract}

\maketitle

\section{Introduction}
Anderson's kinetic exchange interaction \cite{Anderson1959a, Anderson1963a} 
is ubiquitous in magnetic molecules \cite{Hoffmann1975, Kahn1993} 
and insulating materials \cite{Kanamori1959a, Goodenough1963a}.
In particular, the kinetic mechanism has been found as dominant contribution to the exchange interaction in various 
transition metal compounds.
The mechanism has been also often advocated as reason for orbital ordering 
in transition metal oxides with orbitally degenerate metal sites \cite{Kugel1982a}, 
especially, in magnetoresistive manganese oxides \cite{Tokura2000}.  

In all these cases the magnetic orbitals are real and the exchange interaction 
in the case of non-degenerate sites is described by Heisenberg Hamiltonian,
$\hat{H} = -\mathcal{J} \hat{\mathbf{S}}_1 \cdot \hat{\mathbf{S}}_2$.
The kinetic exchange interaction originating from virtual electron transfer
between single-occupied orbitals (s-s) is always antiferromagnetic
\cite{Anderson1959a}:
\begin{eqnarray}
 \mathcal{J}_\textrm{s-s} &=& -\left(\frac{2}{U_{12}} + \frac{2}{U_{21}}\right) t^2,
 \label{Eq:Jss_tm}
\end{eqnarray}
where $t$ is the transfer parameter 
and $U_{ij}$ is the electron promotion energy from site $i$ to site $j$.

On the contrary, the electron delocalization between double-occupied and
single-occupied orbitals (s-d) 
always results in a ferromagnetic contribution (the Goodenough's mechanism \cite{Goodenough1963a}):
\begin{eqnarray}
 \mathcal{J}_\textrm{s-d} 
 &\simeq&
 \frac{2{t}^2}{U_{12}}\frac{J_{\rm H}}{U_{12}} + \frac{2{t}^2}{U_{21}}\frac{J_{\rm H}}{U_{21}},
\label{Eq:Jsd_tm}
\end{eqnarray}
where $J_{\rm H}$ is the Hund's rule coupling constant and $t$ is the transfer integral between corresponding orbitals. 
Given the typical ratio $J_{\rm H}/U \simeq 0.1$ \cite{Anderson1959a, Anderson1963a},
$\mathcal{J}_\textrm{s-d}$ is by one order of magnitude smaller than $\mathcal{J}_\textrm{s-s}$.
Then for comparable electron transfer parameters in s-s and s-d processes, the overall coupling is antiferromagnetic, $\mathcal{J}=\mathcal{J}_\textrm{s-s} + \mathcal{J}_\textrm{s-d} <0$.

A weak ferromagnetic interaction (\ref{Eq:Jsd_tm}) is observed
when the electron transfer between the single-occupied orbitals is negligible or zero,
which is achieved for certain geometries of the exchange bridge
\cite{Anderson1963a, Kanamori1959a, Goodenough1963a}.
Similar ferromagnetic contribution appears also for electron delocalization
between single-occupied and empty orbitals,
as well as in the case of degenerate magnetic orbitals (Kugel-Khomskii model) \cite{Kugel1982a}. 
In all these cases the ferromagnetic kinetic contribution arises in the third order of perturbation theory after $t$ and $J_{\rm H}$.

The kinetic exchange mechanism is equally important in $f$ electron systems 
such as lanthanide and actinide compounds \cite{Mironov2003a, Woodruff2013, Santini2009a, Rau2015}. 
However its realization in these materials is expected to be different from transition metal 
compounds due to a more complex structure of multielectronic states on the metal sites, 
involving complex magnetic orbitals. 
The last are stabilized by strong spin-orbit coupling in lanthanides and actinides giving rise 
to unquenched orbital momentum in their low-lying multiplets \cite{Dieke1967a}, 
which persists in any geometry of their environment. 
Unquenched orbital momentum also occurs in many transition metal complexes and fragments 
when the latter possess cubic \cite{Griffith1961a} or axial \cite{glaser2012} symmetry, 
and it was proved that its effects can persist also under significant deformations 
of the ligand environment \cite{chibotaru2005}. 
Nonetheless, despite numerous examples of strongly anisotropic magnetic materials 
with unquenched orbital momentum on the metal sites, the basic features of kinetic exchange interactions 
in them have not been yet elucidated.

In this work the kinetic exchange interaction for metal sites 
with unquenched orbital momentum is investigated. 
On the example of strongly axial doublet states, we show that 
the paradigm of active magnetic orbitals as always belonging to half-filled ones 
does not hold for strongly anisotropic systems with unquenched orbital momentum on sites. 
In such systems the kinetic exchange interaction between single- and double-occupied orbitals 
is found to be of equal strength with conventional kinetic exchange interaction 
between single-occupied orbitals and can even make the entire interaction ferromagnetic.
Contrary to the Goodenough's mechanism (\ref{Eq:Jsd_tm}), 
the s-d kinetic contribution found here appears already in the second order of 
perturbation theory being of the form (\ref{Eq:Jss_tm}).

\begin{figure*}[bt]
\begin{center}
\begin{tabular}{ll}
(a) & (b) \\
\includegraphics[bb=0 0 312 342, width=5cm]{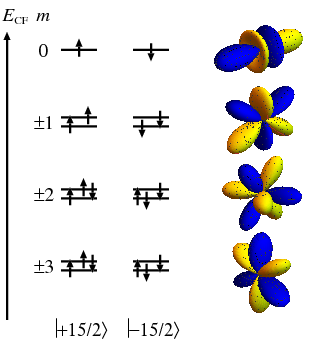}
&
\includegraphics[height=5cm]{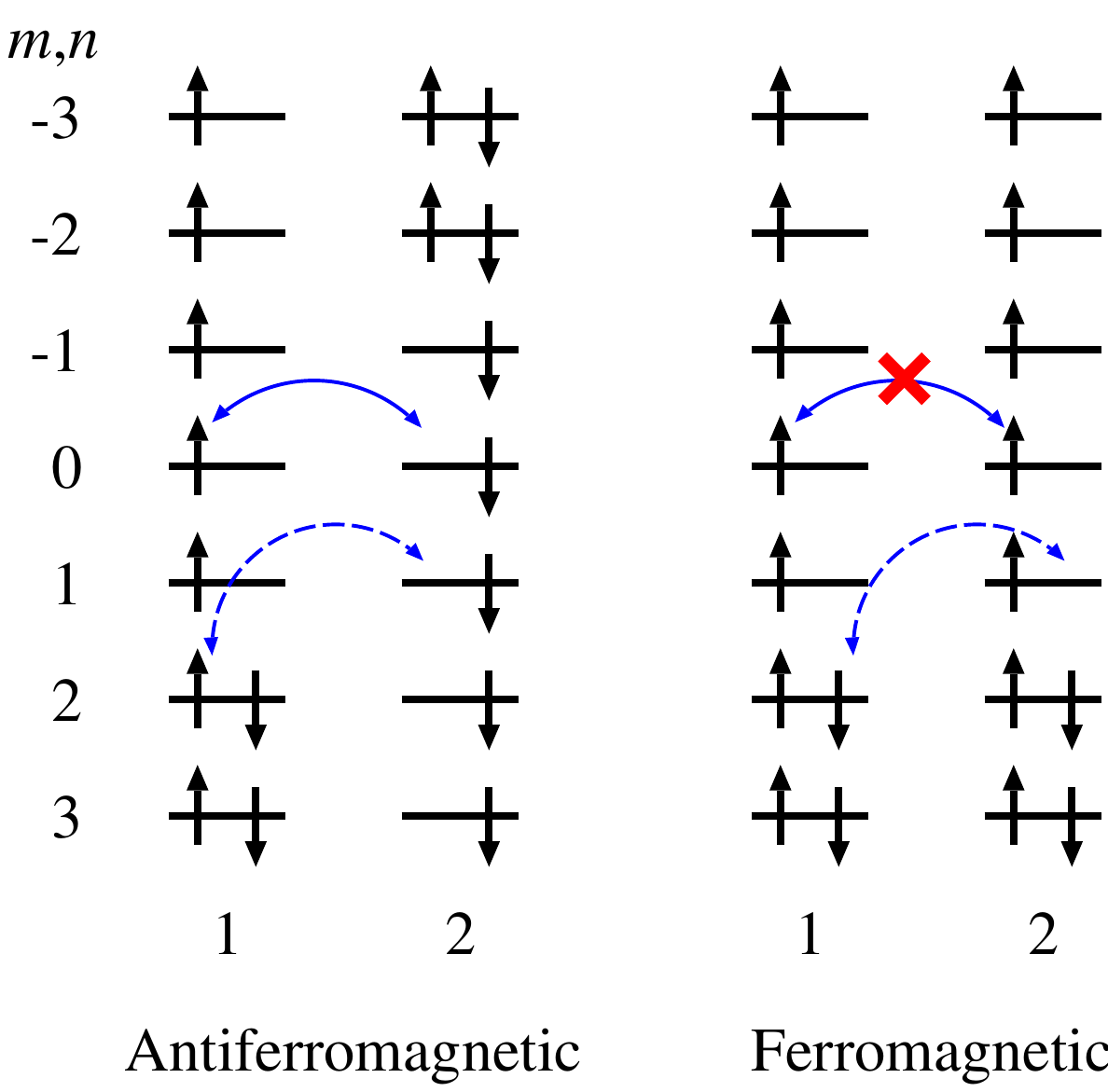}
\end{tabular}
\end{center}
\caption{
Electron transfer processes between doublets with unquenched orbital momentum. 
(a) Scheme of $f$ orbital levels in axial crystal field.
Numbers in the left side stand for the orbital angular momentum projection on the axis of the field. Pictures in the right side show the corresponding real orbitals
(for $m \ne 0$, only one of the two is shown). 
The electron configurations correspond to the wave functions of the Kramers doublet for Dy$^{3+}$ with maximal projection of $J$.
(b) Electron transfer processes between two collinear Dy$^{3+}$ ions in axial Kramers doublets with maximal total momentum projection, $M_J =\pm 15/2$. s-s and s-d processes are shown by solid and dashed lines, respectively. $m, n$ stand for orbital momentum projections on the direction of anisotropy axis on each metal site. Right plots correspond to reversed spin configuration on the site 2.
}
\label{Fig:CF}
\end{figure*}

\section{Results}
\subsection{Doublets with unquenched orbital momentum}
Metal ions are often characterized by non-zero orbital momentum $\hat{\mathbf{L}}$ \cite{Griffith1961a, Abragam1970a}. 
However, in order to keep (part of) it unquenched in complexes and crystals, 
the metal ions should also possess strong spin-orbit coupling 
which splits strongly the atomic (ionic) $LS$ term in multiplets corresponding to 
definite total angular momentum $\hat{\mathbf{J}}=\hat{\mathbf{L}}+\hat{\mathbf{S}}$ \cite{Abragam1970a}. 
This is a standard situation in lanthanides and actinides \cite{Dieke1967a, Santini2009a}. 
Transition metal complexes 
in a threefold degenerate orbital state possess an unquenched orbital momentum corresponding to an effective $\tilde{L}=1$ 
\cite{Griffith1961a}. 
In this case the spin-orbit coupling leads to the formation of multiplets corresponding to total pseudo momentum $\tilde{J}=S+1,\;S,\;|S-1|$.

In low-symmetry crystal field, the (pseudo) $J$-multiplets on metal sites split into Kramers doublets in the case of odd number of electrons, or into singlets in the case of even number of electrons. 
The singlets in the latter case 
form quasi doublets for large $J$ or 
perfectly degenerate (Ising) doublets in environments of axial symmetry \cite{Ungur2011a}. 
In all these doublets the two wave functions are related by time inversion \cite{Abragam1970a}. 
Besides, they are magnetic and contain a significant contribution of orbital momentum 
\cite{comment}. 
The latter necessarily implies that the magnetic orbitals and the wave functions of the doublets are complex. 
In the following we consider the simplest case of an axial crystal field, in which the atomic orbital wave functions preserve the projection of orbital momentum ($\hat{\mathbf{l}}$) on the symmetry axis ($m$). 
The crystal-field orbitals are twofold degenerate with respect to the sign of the projection 
$m$ and are described by the eigenfunctions $|l,\pm m\rangle$ (Fig. \ref{Fig:CF}a). 
For more than half-filled atomic orbital shell $l^N$, $N>2l+1$, 
the ground atomic multiplet corresponds to $J=L+S$ while the wave functions corresponding to the maximal projection, $M_J =\pm J$, are represented by single Slater determinants. 
An example is the ground Kramers doublet of Dy$^{3+}$ ion in strong axial crystal field shown in Fig. \ref{Fig:CF}a \cite{Ungur2011a}.
Further we consider this kind of axial magnetic doublets only \cite{comment3},
which allows us to describe the exchange mechanism in the 
simplest way, though the discussed effects are general for all doublets with unquenched orbital momentum. It is worth mentioning that the doublet states $|J, \pm J\rangle$ appear quite often in the ground state of 
lanthanides and represent a great interest for the design of single-molecule magnets \cite{chibotaru2015}.

\subsection{Exchange interaction for collinear doublets}
The kinetic exchange interaction between doublet states is conveniently described by pseudospin formalism \cite{Abragam1970a}, in which the doublet eigenfunctions $|J, \pm J\rangle$ are put in correspondence to eigenfunctions $|1/2, \pm 1/2\rangle$ of an effective $\tilde{S}=1/2$. 
First, we consider the case of collinear doublets, when their main magnetic axes are parallel.
Since one-electron transfer processes neither can switch nor mix the two doublet wave functions on each metal site,
for relatively large $J$, 
the exchange Hamiltonian reduces to the following Ising form \cite{Chibotaru2015Ising}:
\begin{eqnarray}
 \hat{H} &=& - \mathcal{J} \tilde{S}_{1z} \tilde{S}_{2z},
\label{Eq:Hex}
\end{eqnarray}
where $\tilde{S}_{iz}$ is the $z$ component of the $\tilde{\mathbf{S}}_i$, 
directed along the main magnetic axis on the corresponding metal site 
\cite{comment2}.
In this case the exchange parameter $\mathcal{J}$ is simply derived from the difference 
between energies of antiferromagnetic and ferromagnetic configurations in Fig. \ref{Fig:CF}b, 
$\mathcal{J} = 2(E_\textrm{AF} - E_\textrm{F})$.
We calculated separately
the contributions from s-s and s-d processes (Fig. \ref{Fig:CF}b) 
to $E_\textrm{AF}$ and $E_\textrm{F}$ 
in the second order of perturbation theory after electron transfer.
This yields the following contributions of s-s and s-d processes to the 
exchange coupling constant $\mathcal{J}$:
\begin{eqnarray}
 \mathcal{J}_\textrm{s-s} &=&
 - \left(\frac{2}{U_{12}} + \frac{2}{U_{21}}\right) \sum_{m\in s_1} \sum_{n \in s_2}
 |t_{m,-n}|^2,
\label{Eq:Jss}
\\
 \mathcal{J}_\textrm{s-d} &=&
 - \frac{2}{U_{12}} \sum_{m\in d_1} \sum_{n\in s_2} \left(|t_{m,-n}|^2-|t_{m,n}|^2\right)
\nonumber\\
 &-& \frac{2}{U_{21}} \sum_{m\in s_1} \sum_{n\in d_2} \left(|t_{-m,n}|^2-|t_{m,n}|^2\right),
\label{Eq:Jsd}
\end{eqnarray}
where 
$m$ and $n$ denote the orbitals on site 1 and 2, respectively, by corresponding angular momentum projections (Fig. \ref{Fig:CF}b),
and $s_i$ and $d_i$ indicate the sets of single- and double-occupied orbitals in the electron 
configuration $|J,J\rangle$ of site $i$, respectively. 
For example, for Dy$^{3+}$ ion (site 1 in Fig. \ref{Fig:CF}b) $s_i = \{-3,-2,-1,0,1\}$ and $d_i = \{2,3\}$.
In these equations, $t_{m,n}$ are electron transfer parameters between orbitals $m$ and $n$. 
Note that we do not include effects $\propto J_{\rm H}$ (Goodenough's mechanism)
as being much weaker compared to the s-d contribution found here ({\it vide infra}).

While Eq. (\ref{Eq:Jss}) looks as a standard expression for the s-s kinetic exchange parameter \cite{Anderson1959a, Anderson1963a}, the s-d kinetic contribution, Eq. (\ref{Eq:Jsd}), does not appear for isotropic magnetic systems in this lowest order of the perturbation theory. We can see that it contains electron transfer terms of both signs, {\it i.e.}, antiferromagnetic and ferromagnetic contributions. The terms with $m=0$ in the two brackets of Eq. (\ref{Eq:Jsd}) mutually cancel because of the relation $|t_{0,n}|=|t_{0,-n}|$
\cite{comment1}.
Another evident cancellation occurs for terms with $n=0$. The other pairs of terms in Eq. (\ref{Eq:Jsd}), 
with $m,n \neq 0$, will not cancel each other unless the 
metal-ligand-metal fragment possesses special point symmetry. 
Therefore, for general geometry of exchange-coupled pairs, 
the s-d kinetic exchange is operative and represents a {\em new mechanism} of exchange interaction, 
proper to strongly anisotropic metal ions with unquenched orbital momentum only. 
The peculiarity of this mechanism is that it is of the order $\sim t^2/U$, {\it i.e.}, 
of similar strength as the s-s kinetic exchange, Eq. (\ref{Eq:Jss}). 
However, at variance with 
the s-s kinetic exchange, the s-d exchange can be both 
antiferromagnetic and ferromagnetic as Eq. (\ref{Eq:Jsd}) shows.

Due to time-reversal symmetry the transfer parameters contributing to Eq. (\ref{Eq:Jsd}) 
satisfy the relations $|t_{m,n}|=|t_{-m,-n}|$. 
Using these relations, the total exchange parameter 
$\mathcal{J}=\mathcal{J}_\textrm{s-s} + \mathcal{J}_\textrm{s-d}$ is obtained as
\begin{eqnarray}
 \mathcal{J} &=& - \left(\frac{2}{U_{12}}+\frac{2}{U_{21}}\right) \sum_{m\in s_1} \sum_{n\in s_2} |t_{m,n}|^2.
\label{Eq:J}
\end{eqnarray}
Despite its similar form to $\mathcal{J}_\textrm{s-s}$ in Eq. (\ref{Eq:Jss}), 
the above expression involves different orbitals in the second summation.

%

\begin{figure*}[tb]
\begin{center}
\begin{tabular}{lll}
(a) & (b) & (c) \\ 
\includegraphics[bb=0 0 478 475, width=5cm]{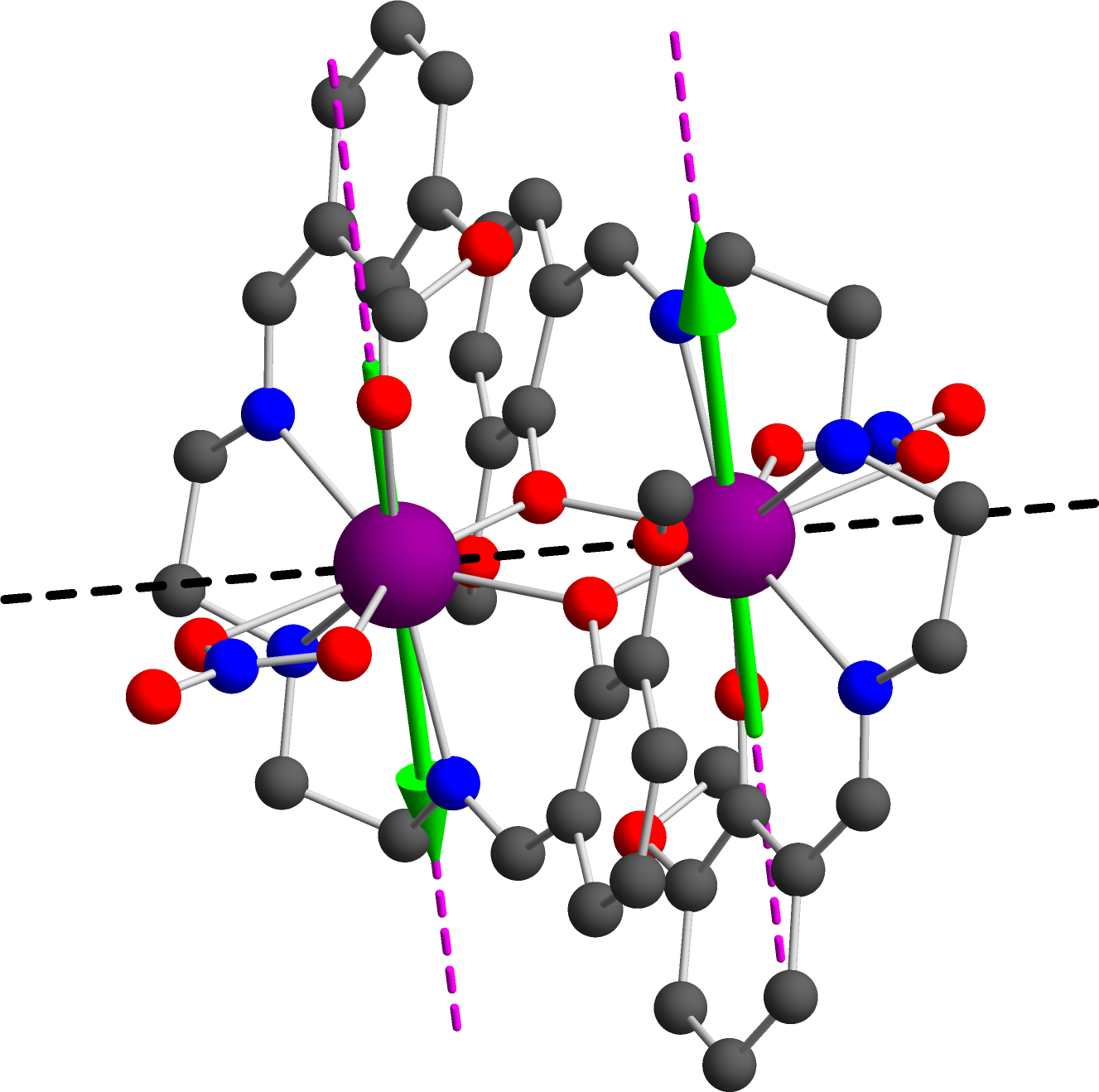}
&
\includegraphics[bb=0 0 417 341, width=5cm]{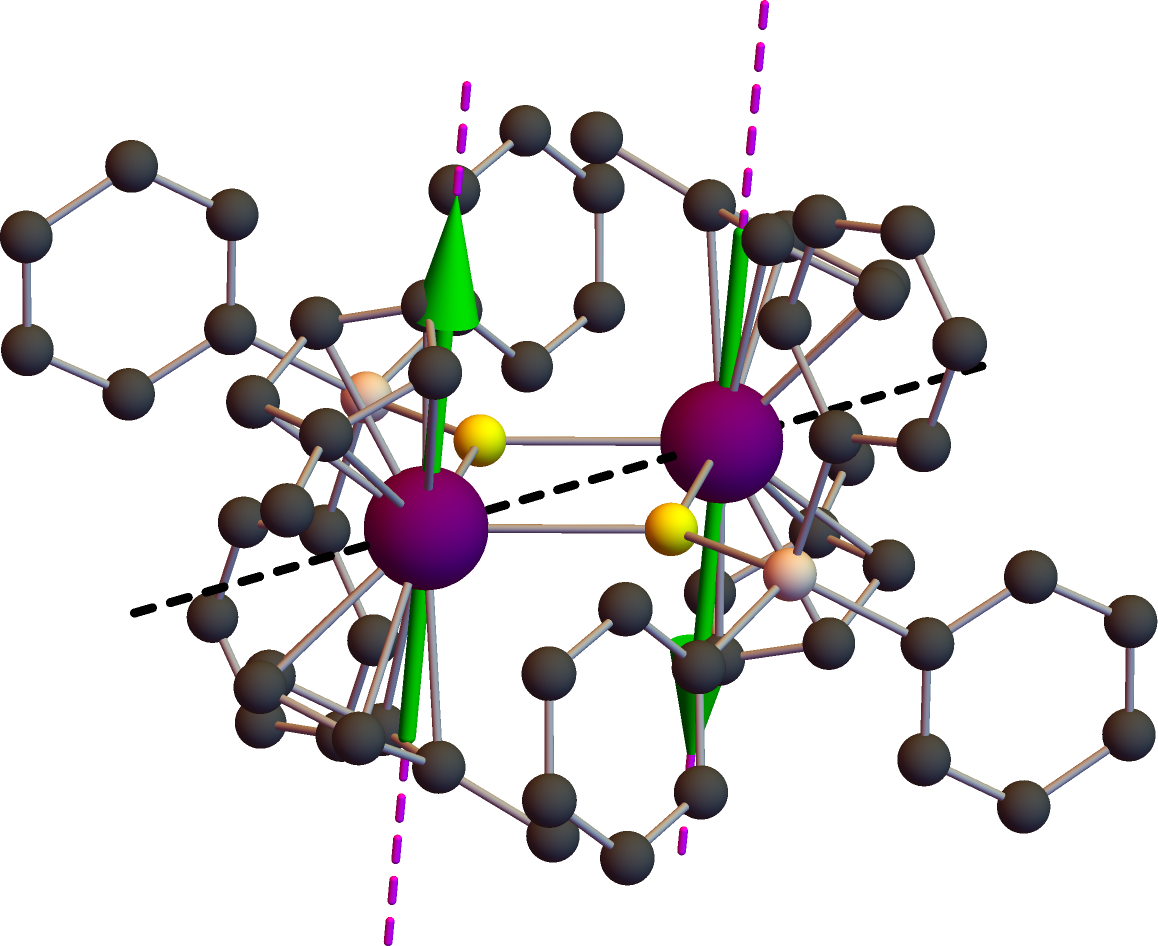}
&
\includegraphics[bb=0 0 455 444, width=5cm]{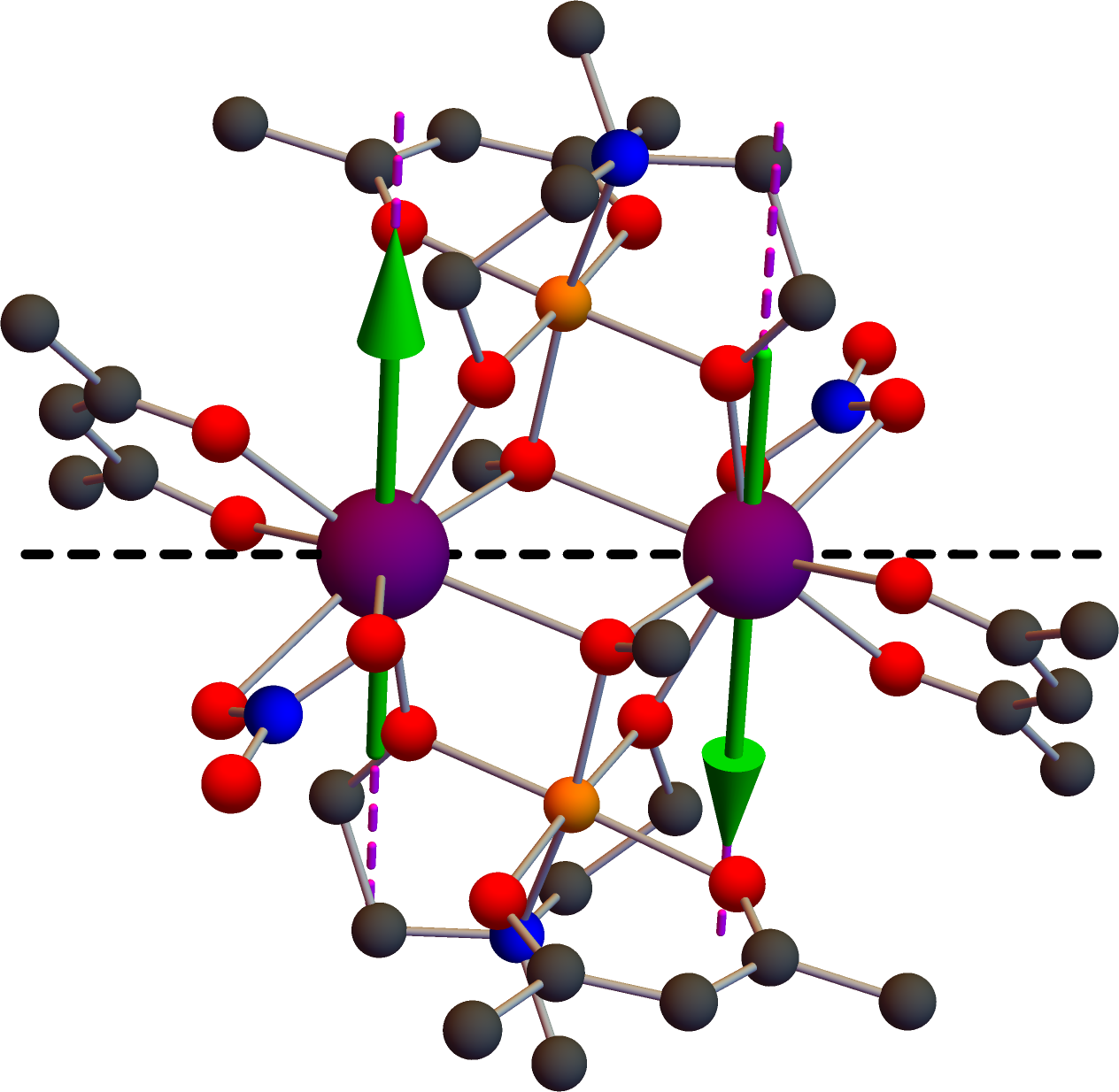}
\end{tabular}
\end{center}
\caption{
The structures of investigated binuclear lanthanide complexes Ln$_2$.
(a) Ln $=$ Tb, Dy, Ho \cite{Long2011}, 
(b) Ln $=$ Dy \cite{Tuna2012}, 
and (c) Ln $=$ Dy \cite{Langley2014}. 
Color legend: Ln purple, O red, C gray, N blue, S yellow, Cr orange, Si beige. 
The pink dashed line is the direction of the main magnetic axis and the green arrow is the magnetic moment
on Ln ions calculated {\it ab initio} \cite{Long2011, Tuna2012, Langley2014}. 
}
\label{Fig2}
\end{figure*}

\begin{table}[tb]
\caption{The exchange coupling parameters $\mathcal{J}$, $\mathcal{J}_\text{s-s}$, 
and $\mathcal{J}_\text{s-d}$ (cm$^{-1}$) for strongly axial magnetic complexes (Fig. \ref{Fig2}).
$\mathcal{J}$ corresponds to experimentally extracted exchange parameter. 
}
\label{Table:J}
\begin{ruledtabular}
\begin{tabular}{cccccc}
System & Ln & Ref. 
& $\mathcal{J}$ & $\mathcal{J}_\text{s-s}$ & $\mathcal{J}_\text{s-d}$ \\ 
\hline
(a) & Tb & \cite{Long2011}\footnote{
The exchange parameter $\mathcal{J}$ was obtained from experimental Ising parameter 
after extracting the magnetic dipole interaction.
}
& $-3.57$ & $-3.58$ & 0.01 \\
(a) & Dy & \cite{Long2011}$^\textrm{a}$    & $-2.97$ & $-2.51$ & $-0.46$ \\ 
(a) & Ho & \cite{Long2011}$^\textrm{a}$    & $-3.22$ & $-1.84$ & $-1.38$ \\
(b) & Dy & \cite{Tuna2012}    & $-2.20$ & $-1.78$ & $-0.42$ \\ 
(c) & Dy & \cite{Langley2014} & $-0.51$ & $-0.51$ &  $0.00$ \\ 
\end{tabular}
\end{ruledtabular}
\end{table}

The kinetic exchange interaction between a strongly axial doublet 
and an isotropic spin is described by the same Ising Hamiltonian (\ref{Eq:Hex}) in which one of the pseudospin operators is replaced by the real spin projection $\hat{S}_z$ of the corresponding site. The expressions for the exchange parameters coincide with Eqs. (\ref{Eq:Jss}) and (\ref{Eq:Jsd}), in which the second summation runs over real orbitals ($a$) for isotropic spin site.
Applying the same argument as for the orbital $m=0$ in the previous case \cite{comment1}, 
we come to the relations $|t_{-n,a}|=|t_{n,a}|$ which cancel the terms in each bracket of Eq. (\ref{Eq:Jsd}). Thus no s-d kinetic mechanism is expected in this case.

{\it Assessment of $\mathcal{J}_\text{s-s}$ and $\mathcal{J}_\text{s-d}$ in lanthanide complexes ---.}
To assess the importance of s-d contribution to the exchange interaction in real complexes, 
we performed a density functional theory (DFT) based analysis of $\mathcal{J}_\text{s-s}$ and $\mathcal{J}_\text{s-d}$
for several previously investigated lanthanide complexes \cite{Long2011, Tuna2012, Langley2014}.
To this end, we first made the localization of Kohn-Sham orbitals on the metal centers and the bridging ligand. 
This allowed us to extract the metal-ligand transfer parameters (the metal-metal ones
turned out to be negligibly small in this approach). 
This tight-binding model together with the Hubbard repulsion energy 
(described by one single parameter $U$ due to the equivalence of the metal sites, see Fig. \ref{Fig2})
was downfolded on the ground spin-orbit doublet states of Ln ions (Fig. \ref{Fig:CF}a). 
This allowed us to calculate straightforwardly the energies of ferromagnetic and antiferromagnetic 
configurations (Fig. \ref{Fig:CF}b) and to obtain the corresponding total $\mathcal{J}$ in Eq. (\ref{Eq:Hex}). 
Then, repeating this procedure by blocking electron transfer processes from double occupied orbitals on the Ln 
sites, we obtain a net s-s contribution to the exchange coupling, $\mathcal{J}_\text{s-s}$, and finally the
s-d contribution: $\mathcal{J}_\text{s-d} = \mathcal{J} - \mathcal{J}_\text{s-s}$. 
In this calculations, the parameter $U$ was chosen to reproduce the experimental exchange parameter $\mathcal{J}$
(Table \ref{Table:J}).

The obtained s-s and s-d contributions are given in Table \ref{Table:J}.
The Dy and Ho complexes from isostructural series (a) and the Dy complex (b) show that the s-d contribution
is by far not negligible in comparison with the s-s contribution.
The increase of the s-d contribution with Ln atomic numbers in the isostructural series (a) is explained 
by the increase of the number of the double-occupied orbitals. 
On the other hand, the vanishing s-d contribution in the complex (c) is due to the cancellation 
of the ferromagnetic and antiferromagnetic contributions in the expression for $\mathcal{J}_\text{s-d}$, Eq. (\ref{Eq:Jsd}).

\subsection{Exchange interaction for non-collinear doublets}
\begin{figure*}[bt]
\begin{center}
\begin{tabular}{ll}
(a) & (b) 
\\
\includegraphics[width=7cm]{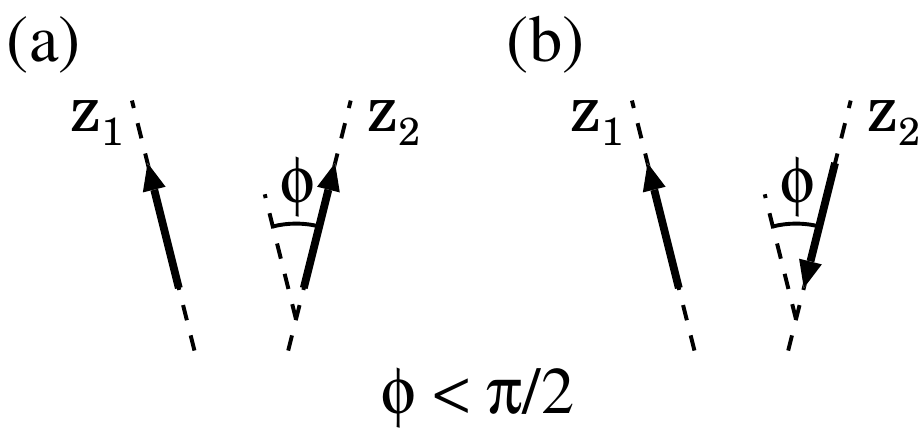}
&
\includegraphics[width=7.0cm]{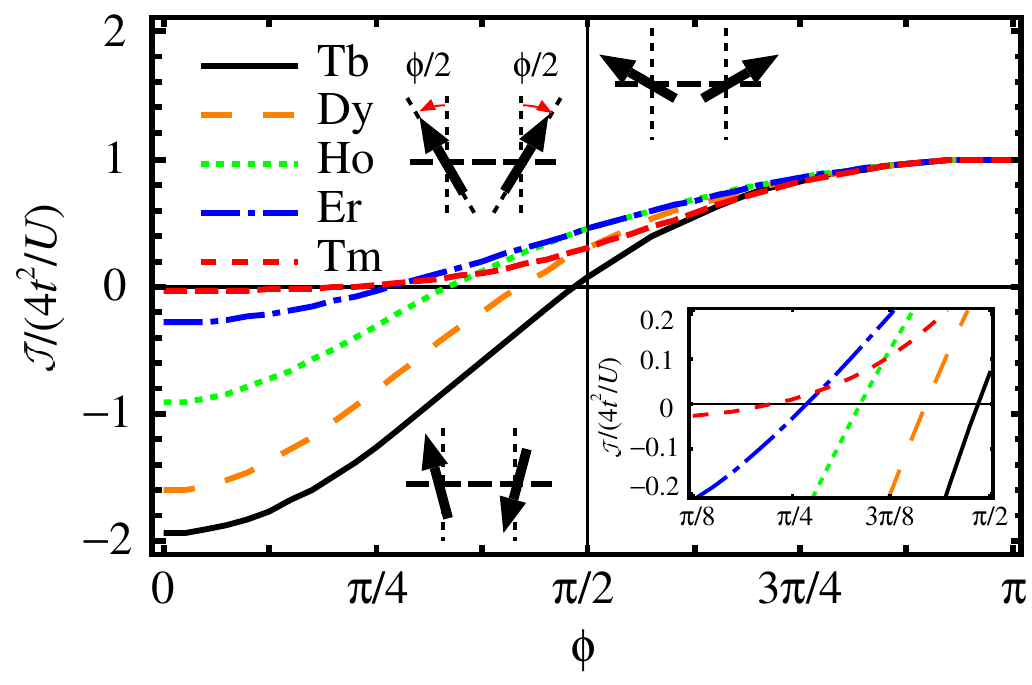}
\\
(c) & (d)\\
\includegraphics[width=7.0cm]{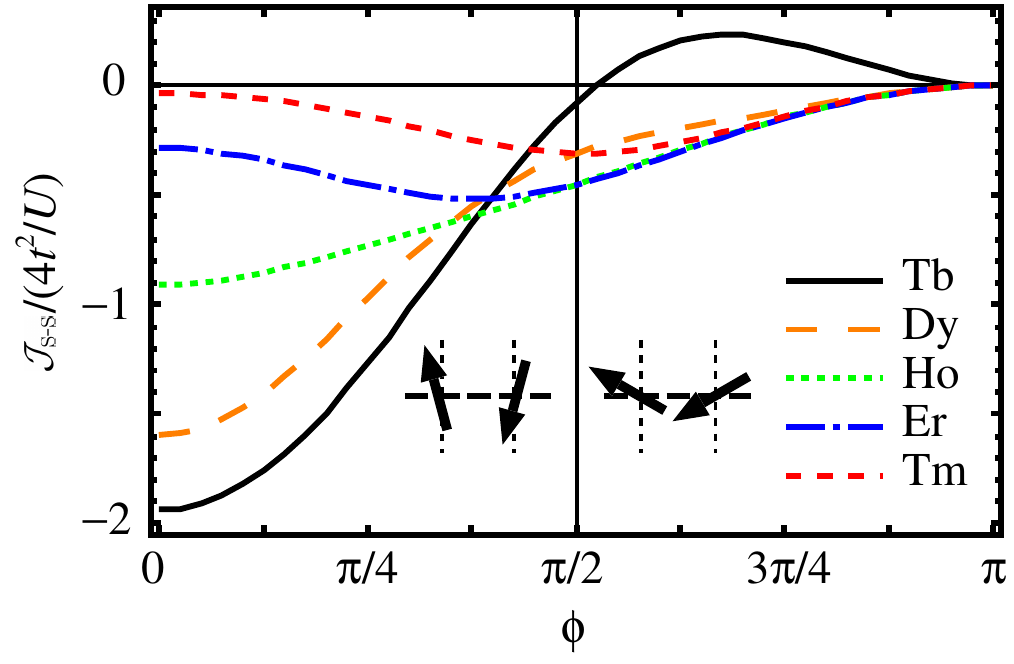}
&
\includegraphics[width=7.0cm]{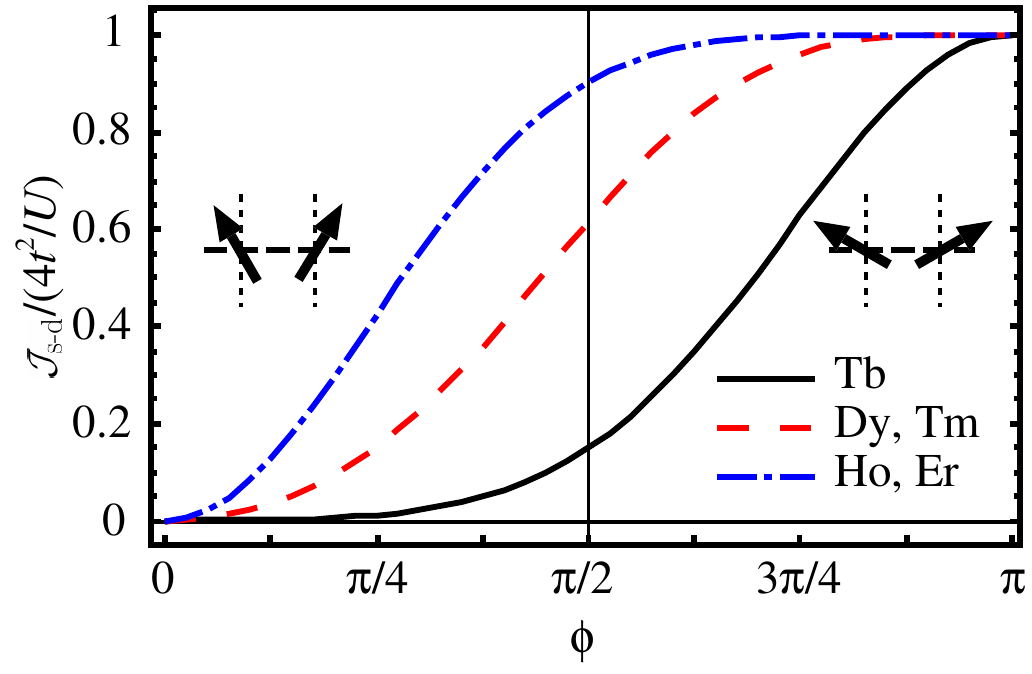}
\end{tabular}
\end{center}
\caption{
s-s and s-d exchange contributions in non-collinear system.
(a) Definition of ferro (i) and antiferro (ii) ordering for non-collinear case. 
The main magnetic axes $z_1$ and $z_2$ are generally non-coplanar.
(b) $\mathcal{J}$, (c) $\mathcal{J}_\textrm{s-s}$, and (d) $\mathcal{J}_\textrm{s-d}$
for Ln$^{3+}$ dimers.
in the symmetric exchange model (inset of plot (a)) 
as function of the angle $\phi$ between the local main magnetic axes. 
The only non-zero transfer parameter is $t\equiv t_{33}^{(0)} = t_{-3-3}^{(0)}$.}
\label{Fig:DeltaE}
\end{figure*}

In non-collinear magnetic systems the main magnetic axes on sites make an angle $\phi$ 
(Fig. \ref{Fig:DeltaE}a). 
The new feature which appears in this case is that electron can transfer to an orbital 
of a neighbor site in both ferro and antiferro configurations (see the definition in Fig. 3), 
with the probability depending on $\phi$. 
As in the collinear case, the single-electron transfer processes cannot switch the multiplet components, 
$|J,J \rangle \rightleftarrows |J,-J \rangle$, when $J$ is sufficiently large \cite{Chibotaru2015Ising}. 
Therefore, the exchange interaction will be described by the same Ising Hamiltonian (\ref{Eq:Hex}) 
with the difference that now pseudospin operators describe momentum projections 
along corresponding main magnetic axes ($z_1$ and $z_2$ in Fig. \ref{Fig:DeltaE}a).   
The exchange parameter corresponding to s-s processes is obtained as
\begin{eqnarray}
 \mathcal{J}_\textrm{s-s} &=&
 -\left(\frac{2}{U_{12}} + \frac{2}{U_{21}}\right) 
\nonumber\\
 &\times& 
 \sum_{m\in s_1} \sum_{n \in s_2}
 \left(\cos^2\frac{\phi}{2} |t_{m,-n}|^2
 - \sin^2\frac{\phi}{2}|t_{m,n}|^2\right),
\label{Eq:Jss_nc}
\end{eqnarray}
and contains now both ferro and antiferro contributions.
On the other hand the exchange parameter for the s-d processes 
remains unchanged, Eq. (\ref{Eq:Jsd}). 
One can see from Eq. (\ref{Eq:Jss_nc}) that $\mathcal{J}_\textrm{s-s}$ is not proportional to $\cos \phi$ unless we have an additional condition $|t_{m,n}| = |t_{m,-n}|$. As was discussed above, the latter is fulfilled for interacting axial doublet and isotropic spin, in which case also the s-d contribution, Eq. (\ref{Eq:Jsd}), vanishes. One should note that the transfer parameters in Eqs. (\ref{Eq:Jsd}) and (\ref{Eq:Jss_nc}) are defined for orbitals quantized along main magnetic axes on the corresponding metal sites and, therefore, are implicitly dependent on angle $\phi$. 

\subsection{Ferromagnetic kinetic exchange interaction}
Contrary to collinear case for which $\mathcal{J}$ is always antiferromagnetic (Eq. (\ref{Eq:J})), the contributions $\mathcal{J}_\textrm{s-s}$ and $\mathcal{J}_\textrm{s-d}$ can be of either sign in non-collinear systems, so that the resulting exchange interaction can be both ferro and antiferromagnetic. 
To investigate this situation we consider a symmetric homonuclear dimer model.
We assume that the electron transfer only takes place between one pair of orbitals,  
$t\equiv t_{\mu \mu}^{(0)} = t_{-\mu -\mu}^{(0)} \ne 0$,
where $\pm\mu$ are orbital momentum projections on the common axis $z$ connecting the metals. 
Figure \ref{Fig:DeltaE}(a) shows calculated $\mathcal{J}$ for $\mu = 3$ as function of $\phi$. 
For small angles, $0 \le \phi \lesssim \pi/4$, $\mathcal{J}$ is always antiferromagnetic. 
In this domain $|\mathcal{J}|$ decreases with increasing $\phi$ and  
at some critical $\phi_c < \pi/2$ becomes ferromagnetic.
Remarkably, the magnitude of the ferromagnetic $\mathcal{J}$ is of the order $\sim t^2 /U$ and its relative strength 
gradually increases when approaching the end of the lanthanide series.
Figures \ref{Fig:DeltaE}(b),(c) show the evolution of $\mathcal{J}_\textrm{s-s}$ (\ref{Eq:Jss_nc}) 
and $\mathcal{J}_\textrm{s-d}$ (\ref{Eq:Jsd}).
For $\phi<\pi/2$,
the s-s and s-d processes stabilize the antiferro and ferromagnetic states, respectively. 
They are found in competition and the latter (s-d) begins to exceed the former (s-s) at a critical $\phi_c$, which has a simple explanation.
The number of single-occupied orbitals decreases with the increase of the number of $f$ electrons ($N$).
Following this trend, the $\mathcal{J}_\textrm{s-s}$ will always decrease with $N$.
On the contrary, $\mathcal{J}_\textrm{s-d}$ roughly depends on the multiplication of the number of single- and double-occupied orbitals. This is the reason why it first increases with $N$ till $N=11$
and then begins to decrease.
As a result, the critical $\phi_c$ decreases with the increase of $N$ and in the cases of Ho, Er and Tm complexes, the critical $\phi_c$ is as small as ca $\pi/4$. The reasons given above explain also the decrease of $\mathcal{J}$ in the domain $0 < \phi < \pi/2$ when moving towards the end of the lanthanide series (Fig. \ref{Fig:DeltaE}(a)).
For $\phi > \pi/2$, the s-s and s-d processes tend to stabilize the ferro and antiferromagnetic states,
respectively. 
In this domain the contribution from the s-d processes is dominant
because the contribution from s-s processes gradually decreases with $\phi$ and becomes completely quenched at $\phi = \pi$.

\begin{figure*}[bt]
 \begin{center}
  \begin{tabular}{ll}
     (a) & (b) 
\\
     \includegraphics[height=7cm, bb=0 0 1291 1139]{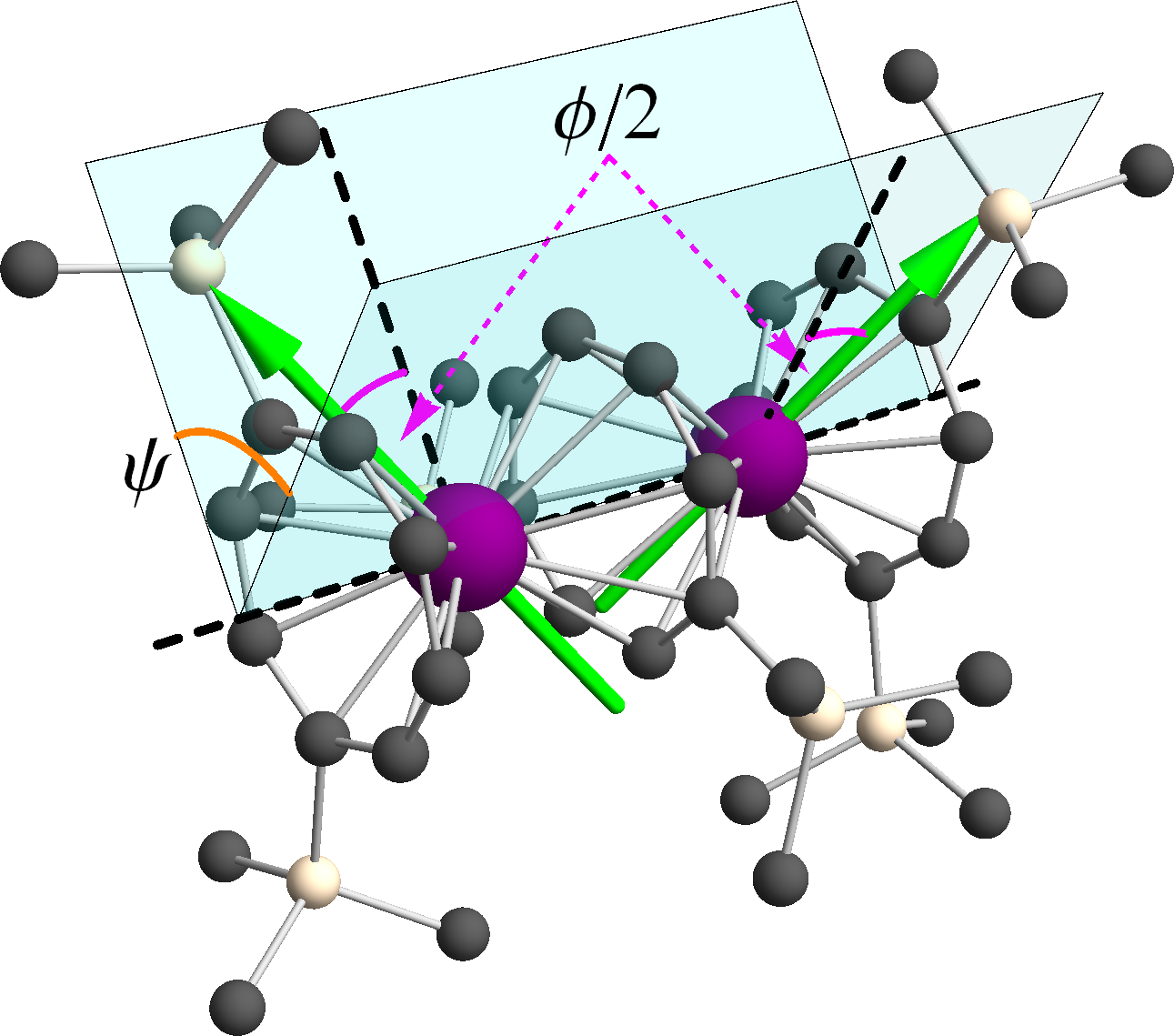}
&
     \includegraphics[height=7cm, bb=0 0 703 616]{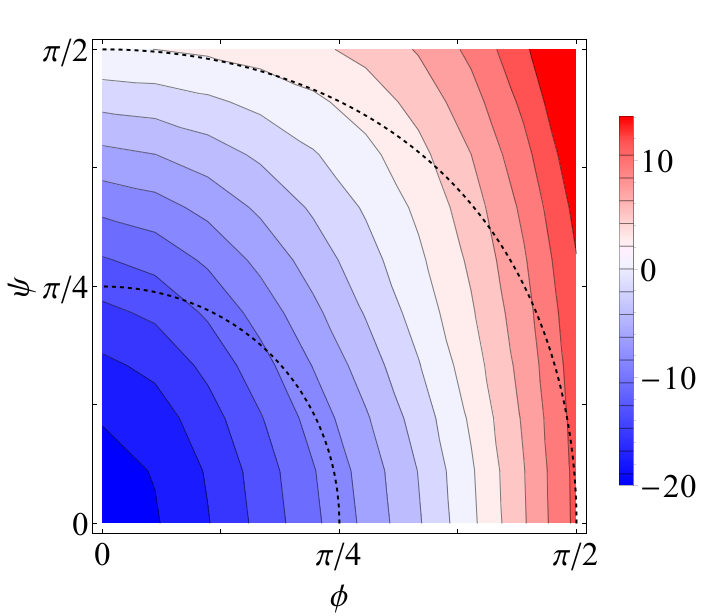}
 \end{tabular}
\end{center}
\caption{
Ferromagnetic exchange interaction induced by strong magnetic anisotropy.
(a) The definition of the angles $\phi$ and $\psi$ defining the relative orientation in the Er$_2$
complex with equivalent metal sites.
(b) Variation of $\mathcal{J}$ with respect to $\phi$ and $\psi$.
The internal dashed line corresponds to the angle $\pi/4$ and the external to the angle $\pi/2$
between the magnetic axes. 
The blue and the red regions stand for the negative and positive values of $\mathcal{J}$, respectively (cm$^{-1}$).
}
\label{Fig4}
\end{figure*}

The change of the sign of kinetic exchange parameter is not specific only to lanthanides.
Similar results are obtained for compounds with transition metal sites in axial ground doublet states 
with unquenched orbital momentum.
We obtain again that in the domain $\pi/4 < \phi < \pi/2$ the exchange parameter becomes ferromagnetic. 
Moreover, for $d^7$ and $d^8$ metal ions this can attain values of $\sim t^2/U$, which corresponds to a very strong ferromagnetic coupling for transition metal compounds
(see Supplemental Materials). 

{\it Effect of non-collinearity on exchange coupling in Er$_2$ complex ---.}
The evolution of the exchange parameter in function of the angle between the magnetic axes 
on metal sites is studied on the example of an Er$_2$ complex \cite{LeRoy2014} (Fig. \ref{Fig4}a).
The calculation has been done in full analogy with the previous case of collinear Ln$_2$
complexes (Fig. \ref{Fig2} and Table \ref{Table:J}).
Similarly to the model calculations (Fig. \ref{Fig:DeltaE}), with the increase of $\phi$
the antiferromagnetic exchange interaction becomes ferromagnetic around $\phi_c \approx 2\pi/5$ (Fig. \ref{Fig4}b).
The shift of $\phi_c$ in comparison with the model calculations 
is due to the existence of many electron transfer processes in this complex. 
In real systems, the direction of main magnetic axes could be controlled by varying ligand environment.

\section{Conclusions}
In this work, we investigated the kinetic exchange interaction
between axial magnetic doublets with unquenched orbital momentum. 
We find a new mechanism of exchange interaction based on electron transfer between single- and double-occupied orbitals. 
Contrary to conventional spin systems, the s-d kinetic contribution found here is not related to 
Goodenough's mechanism (\ref{Eq:Jsd_tm}), arising due to the Hund's rule coupling ($J_{\rm H}$) on metal sites,
but due to the second-order kinetic mechanism (\ref{Eq:Jss_tm}).
On this reason, this kinetic contribution is as strong as the conventional 
kinetic exchange between single-occupied orbitals but, at variance with the latter, can be ferromagnetic. 
In non-collinear magnetic systems the s-d kinetic mechanism can cause an overall ferromagnetic exchange
interaction of the order of $t^2/U$, 
starting from angles $\sim \pi/4$ between main magnetic axes.
These conclusions are fully supported by quantum chemistry based analysis of Ln$_2$ complexes. 
The key feature underlying the new mechanism is that the double-occupied orbitals change 
under time inversion in strongly anisotropic sites due to unquenched orbital momentum.
This is found in sharp contrast to the case of isotropic and weakly anisotropic sites,
where no change of double-occupied orbitals occur under time inversion. 
The obtained results offer a new view on the exchange interaction in lanthanides, actinides and transition metal ions with unquenched orbital momentum. In particular, they show the way to achieve strong ferromagnetic coupling between metal ions, a long sought goal in magnetic materials \cite{Kahn1993}.

\section*{Materials and Methods}
The DFT calculations have been done with the ORCA package \cite{orca}, using 
B3LYP exchange-correlation functional \cite{B3LYP}, in which the Hartree-Fock contribution to the exchange
part was increased from 20 \% to 40 \%. 
This was done to reproduce the experimental isotropic exchange parameters in isostructural Gd$_2$ analogues
of investigated complexes. 
The derivation of tight-binding Hamiltonian for localized Kohn-Sham orbitals and the projection of the Hubbard
model on the ground doublets of investigated Ln$_2$ complexes is described in Supplemental Materials.

\begin{acknowledgments}
We thank V. Vieru for providing us the exchange parameters for the first three complexes in Table \ref{Table:J}.
N. I. would like to acknowledge the financial support from the Flemish Science Foundation (FWO)
and the GOA grant from KU Leuven.
\end{acknowledgments}


%

\clearpage
\begin{center}
\textbf{
Supplemental Materials\\
for
\\
``New mechanism of kinetic exchange interaction induced by strong magnetic anisotropy''
}
\end{center}
\setcounter{equation}{0}
\setcounter{figure}{0}
\setcounter{table}{0}
\setcounter{section}{0}
\makeatletter
\renewcommand{\theequation}{S\arabic{equation}}
\renewcommand{\thefigure}{S\arabic{figure}}
\renewcommand{\thetable}{S\arabic{table}}
\renewcommand{\bibnumfmt}[1]{[S#1]}
\renewcommand{\citenumfont}[1]{S#1}

This material contains:
\\
1) The lanthanide complexes and modeling for the electronic structures; 
\\
2) The density functional theory calculations; 
\\
3) The extraction of the Ising exchange parameters; 
\\
4) The s-s and s-d exchange mechanisms in the $d$ metal complexes.

\section{Model Hamiltonian for binuclear magnetic systems}
As the anisotropic magnetic systems, we treat binuclear lanthanide complexes
\cite{Long2011S, Tuna2012S, Langley2014S, LeRoy2014S}.
In order to describe the low-energy exchange states of these systems, 
the Hubbard Hamiltonian was used:
\begin{eqnarray}
 \hat{H} &=& \hat{H}_0 + \hat{H}_\text{t} + \hat{H}_\text{bi},
\label{Eq:H}
\\
 \hat{H}_0 &=& \sum_{r \sigma'} \epsilon_r \hat{n}_{r\sigma'}, 
\label{Eq:H0}
\\
 \hat{H}_\text{t} &=& \sum_{i=1,2} \sum_{m\sigma} \sum_{r \sigma'} 
 \left( \tau^{iL}_{mr} D_{\sigma \sigma'}^\frac{1}{2} (R_i)
 \hat{c}_{im\sigma}^\dagger \hat{c}_{Lr\sigma'}
 \right.
\nonumber\\
 &+& 
 \left.
 \tau^{Li}_{rm} D_{\sigma \sigma'}^{\frac{1}{2}*} (R_i)
 \hat{c}_{Lr\sigma'}^\dagger \hat{c}_{im\sigma}
 \right),
\label{Eq:Ht}
\\
 \hat{H}_\text{bi} &=& \sum_{i=1,2} \sum_{m\sigma \ne  m'\sigma'} \frac{u_f}{2} \hat{n}_{im\sigma} \hat{n}_{im'\sigma'}.
\label{Eq:Hbi}
\end{eqnarray}
Here, $\hat{H}_0$, $\hat{H}_\text{t}$, $\hat{H}_\text{bi}$ indicate 
the orbital energy level of the bridging atoms with respect to the $f$ orbital level, 
the electron transfer Hamiltonian between metal site and the bridging atoms (ligand), 
and $\hat{H}_\text{bi}$ is the Coulomb repulsion on $f$ metal sites, respectively.
In the Hamiltonian, $i$ $(= 1,2)$ is the metal (lanthanide) center, $L$ is the bridging ligand,
$m$ $(= 3, 2, ..., -3)$ is the projection of the orbital angular momentum of $f$ atomic orbital,
$r$ is the orbital energy level of the ligand, 
$\sigma, \sigma'$ is the spin projection, 
$\hat{c}_{im\sigma}$ ($\hat{c}_{im\sigma}^\dagger$) and 
$\hat{c}_{Lr\sigma}$ ($\hat{c}_{Lr\sigma}^\dagger$) are the
annihilation (creation) operators on the $m\sigma$ orbital of metal site and $r\sigma$ orbital of the ligand $L$, 
respectively, 
$\hat{n}_{im\sigma} = \hat{c}_{im\sigma}^\dagger \hat{c}_{im\sigma}$ and 
$\hat{n}_{Lr\sigma} = \hat{c}_{Lr\sigma}^\dagger \hat{c}_{Lr\sigma}$
are the number operators, 
$\tau^{iL}_{mr}$ $(= \tau^{Li*}_{rm})$ is the transfer parameter between magnetic center and ligand, 
and $u_f$ is the Coulomb repulsion energy on $f$ metal site.
Since the transfer parameter between metal sites are small, we neglect the direct electron transfer 
between the metal centers.
The Coulomb repulsion on the ligand and the intersite one between the metal center and the ligand
as well as the Hund's rule coupling are neglected because they are smaller than $u_f$.
Based on the Hamiltonian, we describe the low-energy magnetic states. 
The transfer parameters are extracted from density functional theory (DFT) calculations and 
the Coulomb repulsions are determined to reproduce the experimental exchange parameters. 

The model Hamiltonian Eq. (\ref{Eq:H}) is not the same as the model Hamiltonian 
used for the derivation of the exchange interactions in the main text. 
In the latter model Hamiltonian, the electron transfer between the metal sites and ligand is reduced to the
transfer parameter between the metals.
However, in the DFT based calculations of the exchange parameters, 
such simplification is not always possible because 
the $f$ orbital levels and ligand levels are sometimes close to each other.
Thus, we take Eq. (\ref{Eq:Ht}) for our calculations of the lanthanide complexes.

\section{DFT calculations}
In order to find a density functional which is suitable 
for the electronic structure calculations of lanthanide complexes, 
we calculated (i) the exchange interactions and (ii) the binding energies with
several methods (DFT and Hartree-Fock (HF) method), and compared the results with experimental data.
We found that the hybrid functional including about 40 \% of HF exchange 
is suitable to express one-orbital parameters 
($\epsilon_r$ and $\tau^{iL}_{mr}$ in Eqs. (\ref{Eq:H0}) and (\ref{Eq:Ht})) of lanthanide complexes.
For the quantum chemistry calculations, we used ORCA3.0 package \cite{orcaS}.

\subsection{Isotropic exchange parameters for Gd$^{3+}$ complexes}
In general, the spin-orbit coupled ground states of the lanthanide ion (Dy$^{3+}$, Er$^{3+}$, etc.) 
cannot be adequately treated within single Slater determinant approach.
On the other hand, the ground state of half-filled Gd$^{3+}$ ion is in a good approximation 
described by pure spin state, and the state with maximal spin projection is a single Slater determinant. 
Thus, for the quantum chemistry calculations, 
we replaced the lanthanide ions (Dy$^{3+}$ and Er$^{3+}$ ions) by Gd$^{3+}$ ions.
As the functional, we chose B3LYP with various HF exchange contributions (20 - 60 \%).
The variation of the contribution of the HF exchange is to reduce the exaggerated 
electron transfer and to increase the underestimated Coulomb repulsion on site. 
The molecular structures were taken from the x-ray diffraction data. As the basis set, SVP was used.

The exchange parameter between metal sites was obtained within broken-symmetry approach \cite{Soda2000S}.
The exchange parameter $\mathcal{J}$ defined by Heisenberg Hamiltonian, 
\begin{eqnarray}
 \hat{H} &=& -\mathcal{J} \hat{\mathbf{S}}_1 \cdot \hat{\mathbf{S}}_2,
\label{Eq:Heisenberg}
\end{eqnarray}
is estimated using the high-spin and broken-symmetry (low-spin) states: 
\begin{eqnarray}
 \mathcal{J} &=& -2\frac{(E_\text{HS} - E_\text{BS})}
                 {\langle \hat{\mathbf{S}}^2 \rangle_\text{HS}- \langle \hat{\mathbf{S}}^2\rangle_\text{BS}}.
\label{Eq:J_Heisenberg}
\end{eqnarray}
Here, $E_\text{HS}$ and $E_\text{BS}$ are the ground electronic energies obtained 
from the high-spin (HS) and broken-symmetry (BS) calculations, 
and $\langle \hat{\mathbf{S}}^2 \rangle_\text{HS}$ and $\langle \hat{\mathbf{S}}^2\rangle_\text{BS}$
are the expectation values of the magnitude of the total spin. 

The obtained isotropic exchange parameters and those from experimental data
are tabulated in Table \ref{Table:BSDFT}.
We find B3LYP functionals with 40-50 \% of Hartree-Fock exchange well reproduce the the experimental ones.

\begin{table*}[tb]
\caption{
The exchange parameters extracted from the broken-symmetry DFT calculations and experimental magnetic susceptibility (cm$^{-1}$).
The numbers in the first row indicate the contribution of the HF part in the exchange-correlation functional. 
The last column are the experimental data (Exp.) extracted using the Lines model or Heisenberg model. 
The experimental data of the Gd complexes (a), (b), (d) are taken from 
Refs. \onlinecite{Long2011S, Tuna2012S, LeRoy2013S}, respectively.
The effect of the magnetic dipolar interaction is removed only for the complex (a).
}
\label{Table:BSDFT}
\begin{ruledtabular}
\begin{tabular}{ccccccccccc}
     & 20 \%  & 25 \%  & 30 \%  & 35 \%  & 40 \%  & 45 \%  & 50 \%  & 55 \%  & 60 \%  & Exp. \\
\hline
(a)  & -0.300 & -0.268 & -0.235 & -0.213 & -0.194 & -0.180 & -0.160 & -0.150 & -0.133 & -0.17 \cite{Long2011S} \\
(b)  & -0.319 & -0.280 & -0.249 & -0.231 & -0.207 & -0.190 & -0.173 & -0.157 & -0.142 & -0.210 \cite{Tuna2012S} \\
(d)  & -1.468 & -1.057 & -0.785 & -0.603 & -0.478 & -0.385 & -0.312 & -0.260 & -0.222 & -0.448 \cite{LeRoy2013S}\\
\end{tabular}
\end{ruledtabular}
\end{table*}

\subsection{Binding energy vs. photoemission measurements}
The binding energies of tris-cyclopentadienyl lutetium (LuCp$_3$) were estimated and compared with experimental data.
The experimental binding energies are taken from the photoelectron spectra (PES) of LuCp$_3$ in gas phase \cite{Coreno2010S}.
We chose Lu$^{3+}$ ion complex because the $4f$ orbitals are completely filled in the ground state, 
which can be described by singe Slater determinant.
For the calculations, we used the x-ray diffraction structure of CeCp$_3$ \cite{Baisch2006S}
as the structure of LuCp$_3$ since their structures are similar to each other
and the energy scale of the PES is much larger than the change in energy due to the structure. 
The binding energies were estimated applying the Koopmans' theorem. 

The calculated binding energies are shown in Table \ref{Table:EB}. 
In comparison with the experimental data, 
the B3LYP (20 \%) and the HF calculations underestimates and overestimates the $4f$ binding energies, respectively.
Better agreement is obtained when the contribution of the HF exchange is about 40 \%.
Since both the exchange interaction and the binding energies obtained with the hybrid B3LYP functional with about 
40 \% of the HF exchange are close to the experimental data, 
we used the functional for the calculations of the transfer Hamiltonian.

\begin{table*}[tbh]
\caption{
Binding energies (eV).
$4f$, Cp $(\sigma)$, and Cp $(\pi)$ are the nature of the Kohn-Sham orbitals for the corresponding range of energies. 
The experimental energies indicate the peaks of the photoelectron spectra \cite{Coreno2010S}.
}
\label{Table:EB}
\begin{ruledtabular}
\begin{tabular}{ccccccc}
     & \multicolumn{6}{c}{B3LYP}\\
     & 20 \% & 25 \% & 30 \% & 35 \% & 40 \% & 45 \% \\
\hline
$4f$          & 12.47 - 12.59 & 13.21 - 13.33 & 13.96 - 14.08 & 14.71 - 14.82 & 15.46 - 15.58 & 16.21 - 16.33 \\
Cp $(\sigma)$ &  9.97 - 11.22 & 10.26 - 11.59 & 10.56 - 11.95 & 10.85 - 12.32 & 11.14 - 12.68 & 11.43 - 13.05 \\
Cp $(\pi)$    &  5.24 -  6.71 &  5.42 -  6.92 &  5.60 -  7.12 &  5.78 -  7.33 &  5.97 -  7.54 &  6.15 -  7.74 \\
\hline
     & \multicolumn{3}{c}{B3LYP} & HF & Exp. \\
     & 50 \% & 55 \% & 60 \% &    & \\
\hline
$4f$          & 16.97 - 17.09 & 18.04 - 17.68 & 18.63 - 18.54 & 23.29 - 23.21 & 14.30, 15.74 \\
Cp $(\sigma)$ & 11.73 - 13.42 & 12.02 - 13.79 & 12.31 - 18.35 & 15.92 - 13.77 & 12.4 \\  
Cp $(\pi)$    &  6.33 -  7.95 &  6.52 -  8.16 &  6.70 -  8.37 &  9.13 -  7.22 & 7.28, 8.81 \\
\end{tabular}
\end{ruledtabular}
\end{table*}

\section{Derivation of Ising exchange parameters for Ln$^{3+}$ complexes}
In order to project the electronic states from the DFT calculations into the model Hamiltonian (\ref{Eq:H}), 
we localized the valence Kohn-Sham orbitals
(Pipek-Mezey localization \cite{Pipek1989S}).  
The Kohn-Sham orbitals are divided them into the magnetic $4f$ orbitals, the bridging ligand orbitals, and the other ligand orbitals,
where the last orbitals are not important for the exchange interaction from each other.
The localization on the magnetic core was measured by the Mulliken population on the 
$f$ type orbitals of the metal centers and the $p$ type orbitals of the bridging ligand.
With the use of the localized orbitals as the basis set, 
we expressed the Kohn-Sham Hamiltonian matrix $\mathbf{h}_\text{KS}$ for the magnetic core.
Here, the localized $f$ orbitals were transformed into the eigenstates of 
the atomic orbital angular momentum $\hat{l}_z$ whose quantization axes agree with the 
{\it ab initio} main magnetic axes \cite{Long2011S, Tuna2012S, Langley2014S}, $|im\sigma\rangle$.
On the other hand, the ligand orbitals $|Lr\sigma\rangle$ are chosen so that 
the bridging ligand part of $\mathbf{h}_\text{KS}$ becomes diagonal. 
Then, the diagonal elements of the Hamiltonian matrix $\mathbf{h}_\text{KS}$ are 
used as the orbital energy levels (the $f$ orbital level is averaged) in Eq. (\ref{Eq:H0}),
and the elements in the off-diagonal block matrices between the metals and 
the bridging ligand as the transfer parameters in Eq. (\ref{Eq:Ht}). 

The matrix of model Hamiltonian (\ref{Eq:H}) was calculated using the 
ground ferromagnetic, the ground antiferromagnetic electron configurations (Fig. 1b in the main text),
and the configurations with one-electron transfer between metal sites and between metal and ligand as the basis set.
Diagonalizing the Hamiltonian matrix, we obtain the ferromagnetic and antiferromagnetic states. 
When there is no energy level between these two states, 
we can project these states into Ising Hamiltonian (Eq. (3) in the main text).
The Coulomb repulsion $u_f$ is determined to reproduce the experimental exchange parameter. 


The obtained exchange parameters are shown in Table I in the main text. 
The Coulomb repulsion energies are 1.50, 1.40, 1.20 eV for the Tb, Dy, Ho complexes in the series (a), 
1.48 and 1.44 eV for the complexes (b) and (c), respectively. 
In all cases, there is no level between the ground ferro- and antiferromagnetic levels.
At a glance, $u_f$ looks too small (cf. Ref. \onlinecite{vanderMarel1988S}), 
whereas it is not too small because the effect of the Coulomb repulsion is partly 
included in the Kohn-Sham orbital energies within the mean-field approximation.
Thus, $u_f$ is the difference between the Hubbard $U$ and the mean-field value. 
The ferromagnetic and the antiferromagnetic ground states are mainly contributed by the electron configurations
without electron transfer (the type of Fig. 1b in the main text).
Their contributions (probabilities) to the ground state are 
98.3 \%, 99.4 \%, 98.9 \% for the Tb, Dy, Ho complexes of the series (a), 
99.3 \% for the complex (b), 99.7 \% for the complex (c), and 97 \% for the Er complex. 
These high probabilities guarantee the validity of the description of the low-lying states by the pseudospin Hamiltonian.

\section{s-s and s-d exchange mechanisms in $d$ metal complexes}
\begin{figure*}[bth]
\begin{center}
\begin{tabular}{lll}
(a)& ~~~~& (b)\\
\includegraphics[height=5cm]{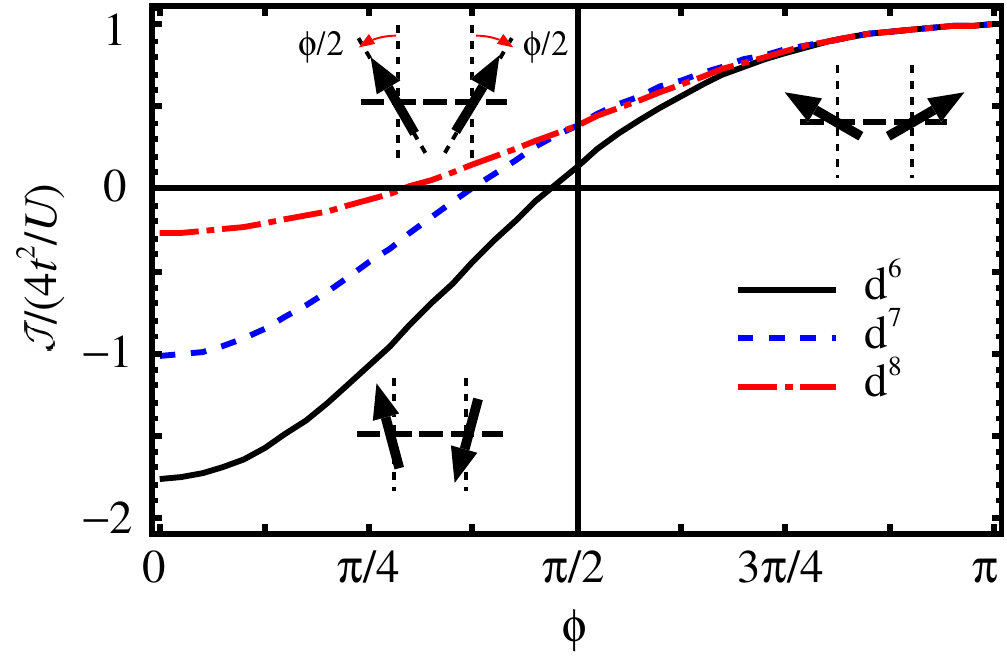}
&
&
\includegraphics[height=5.0cm, bb = 0 0 1196 1050]{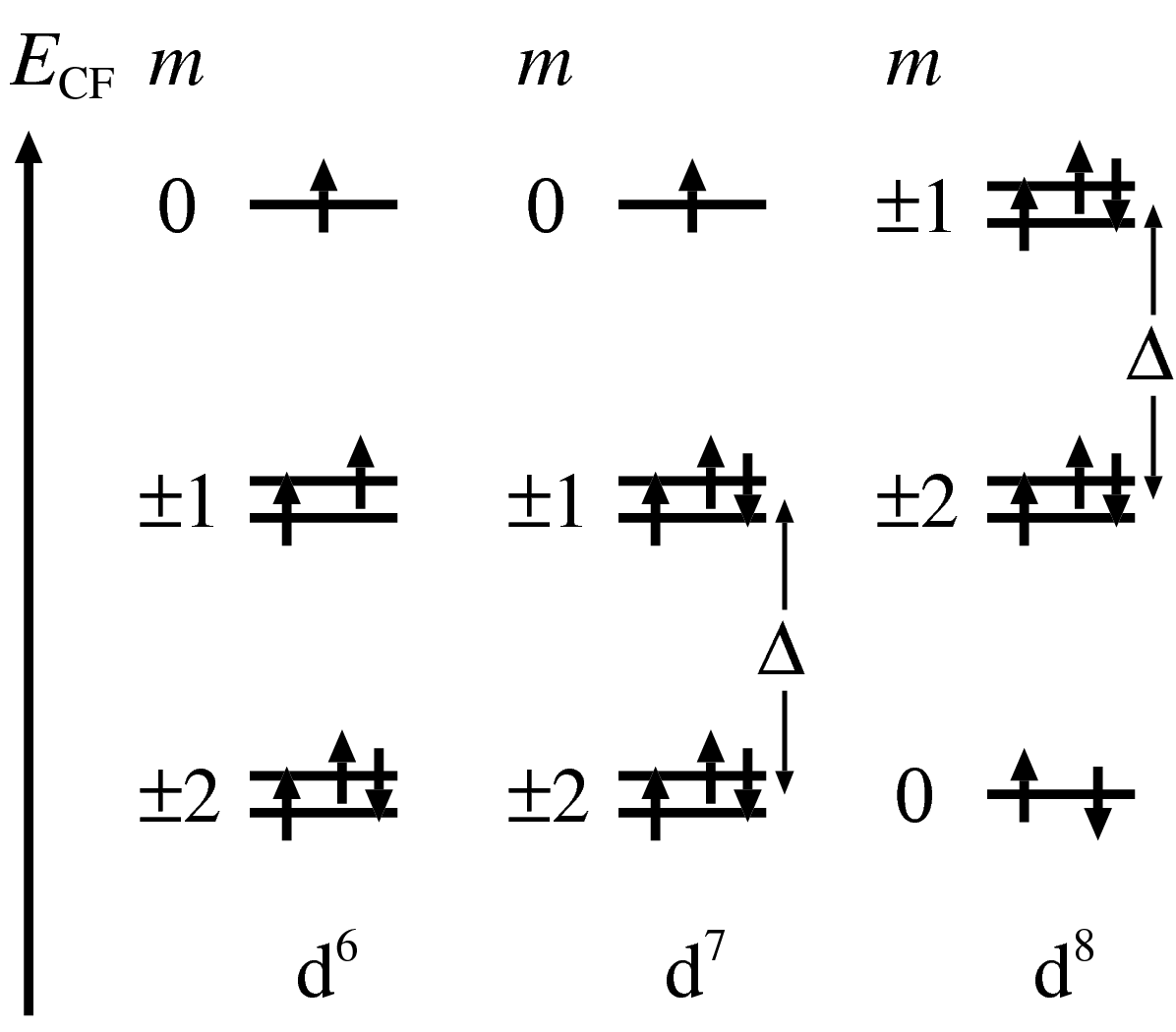}
\end{tabular}
\end{center}
\caption{
(a) $\mathcal{J}$ for several $d$-metal complexes as function of the angle between local main magnetic axes.
(b) $d^6$, $d^7$ and $d^8$ electron configurations corresponding to axial doublets with maximal $M_J$. 
The non-aufbau population in the last two electron configurations is achieved for relatively small separation $\Delta$ between orbital levels with non-zero $m$ (compared to spin-orbit coupling).
This order of orbitals is realized, for example, in crystal fields 
with axial symmetry ($d^6$, $d^7$) and/or trigonal symmetry ($d^8$) \cite{Ungur2013S}.
}
\label{Fig:DeltaE_d}
\end{figure*}

Within the simple two-sites model for the non-collinear doublets in the main text, 
we calculated the exchange interaction parameter for $d$ metal complexes assuming that the ground state is $|J,\pm J\rangle$. 
The exchange parameter $\mathcal{J}$ for the $d^6$, $d^7$ and $d^8$ ions are shown in Fig. \ref{Fig:DeltaE_d}(a).
As in the case of the $f$ metal ions, the exchange becomes ferromagnetic as the increase of angle $\phi$. 

Although the crystal field in transition metal complex is stronger than in lanthanide complex,
the crystal field level with unquenched orbital can be obtained with suitable symmetry of the ligands.
For example, axial ($d^6, d^7$) or trigonal ($d^8$) crystal field splits the $d$ levels into two doublets 
and one nondegenerate state \cite{Ungur2013S}. 
With the splitting, the ground doublets of the $d^6$ metal originates from 
the $J$-multiplet due to the strong Hund's rule coupling (Fig. \ref{Fig:DeltaE_d}(b)).
In the case of the $d^7$ and $d^8$ systems, the orbital momentum is unquenched 
in the presence of the spin-orbit coupling which exceeds the crystal field splitting of the 
orbital levels ($\Delta$ in Fig. \ref{Fig:DeltaE_d}(b)),
which could be observed even in $3d$ metals systems.

\begin{thebibliography}{34}
\bibitem [{com({\natexlab{a}})}]{comment}%
  \BibitemOpen
  \href@noop {} {}\bibinfo {note} {Although no orbital
  momentum is conserved in these doublets, their $J$-multiplet genealogy
  implies large orbital contribution to the total magnetic moment.}\BibitemShut
  {Stop}%
\bibitem [{com({\natexlab{d}})}]{comment1}%
  \BibitemOpen
  \href@noop {} {}\bibinfo {note} {This is due to the real
  orbital corresponding to $m=0$, for which we have $t_{0,n}^* =(-1)^n
  t_{0,-n}$.}\BibitemShut {Stop}%
\end{thebibliography}

%

\end{document}